\documentclass[twocolumn,prl,showpacs,floatfix]{revtex4}
\usepackage{graphicx}
\usepackage{dcolumn}
\usepackage{bm}
\usepackage{amsmath}

\begin{document}

\title{Symmetry-conserving vortex clusters in small rotating clouds of ultracold bosons}

\author{Igor Romanovsky}
\author{Constantine Yannouleas}
%\email{Constantine.Yannouleas@physics.gatech.edu}
\author{Uzi Landman}
%\email{Uzi.Landman@physics.gatech.edu}

\affiliation{School of Physics, Georgia Institute of Technology,
             Atlanta, Georgia 30332-0430}

\date{21 February 2008; Physical Review A {\bf 78}, 011606 (2008), Rapid Communication}

\begin{abstract}
The properties of a special class of correlated many-body wave functions, 
named rotating vortex clusters (RVCs), that preserve the total angular 
momentum of a small cloud of trapped rotating bosons are investigated. 
They have lower energy and provide a superior description for the formation of 
vortices compared to the mean-field Gross-Pitaevskii (GP) states that break the 
rotational symmetry. 
The GP vortex states are shown to be wave packets composed of such RVC states. 
Our results suggest that, for a small number of bosons, the physics is 
different from that of ideal Bose-Einstein condensates which
characterize larger assemblies. 
\end{abstract}

\pacs{03.75.Hh, 03.75.Lm, 03.75.Nt}

\maketitle

Mean-field descriptions of the many-body problem exhibit a ubiquitous
symmetry-breaking behavior that extends across several fields of physics,
from nuclear physics \cite{rsbook}, quantum chemistry \cite{carb90}, and
metallic microclusters \cite{ylbook99} to semiconductor quantum dots 
\cite{yl07} and trapped ultracold atoms \cite{yl07,roma04,muel06}. 

Mean-field broken-symmetry solutions are expected to play the role of an 
effective ground state in the thermodynamic limit, $N \rightarrow \infty$, 
when quantum fluctuations about the mean-field state may be omitted 
\cite{pwa84}. A prominent example of such broken-symmetry states is given
by the Gross-Pitaevskii (GP) vortex states in a harmonic trap \cite{gpbook03}. 
Specifically, although each individual vortex carries a quantized 
amount of angular momentum \cite{baym05}, the GP vortex solutions as a whole  
break the rotational symmetry of the confining harmonic trap \cite{butt99}, 
and thus they are not eigenstates of the total angular momentum $\hat{L}=
\sum_{i=1}^N \hat{l}_i$. In agreement with the general ideas of Ref.\ 
\cite{pwa84}, GP vortex states have been observed experimentally (see, e.g., 
Refs.\ \cite{dali00,kett01,corn01,hodb02}) for rotating Bose gases 
with large $N$. Indeed the energy advantage of symmetry restored states 
(see below) over the mean-field solutions diminishes as $N$ 
increases (see section 1.2. in Ref.\ \cite{yl07}).

Recently, the availability of optical lattices \cite{grei02,lewe07} 
with a small number of particles per lattice site serves to motivate studies 
of small clouds of rotating bosons. For a small number $N$ of atoms, however, 
quantum fluctuations cannot be neglected, and one needs to consider methods  
beyond the mean-field approximation \cite{yl07}. 

A natural way for accounting for quantum correlations about the GP vortex 
solutions in a harmonic trap is the method of restoration of rotational symmetry 
via projection techniques. This method \cite{yl07} was introduced recently
in quantum dots \cite{yl02.1,yl02.2} and 
harmonic traps \cite{roma04,roma06} to describe {\it individual particle 
localization\/} and formation of rotating electron molecules (REMs) \cite{note1}
and rotating boson molecules (RBMs), respectively. Here, we use 
projection techniques to define and study vortex states with {\it good\/} 
total angular momentum, showing that these states can be 
properly referred to as rotating vortex clusters (RVCs).
We stress again that the RVCs are eigenstates of the total angular
momentum in contrast to the GP vortex states. 
For small $N$, the rotating-vortex-cluster states are the natural entities 
to be employed (in place of the GP vortices) for comparisons with exact 
solutions, which are eigenstates of the total angular momentum by their very 
nature. Furthermore we note the generality of the 
methodology of symmetry restoration. It applies as well to other broken symmetries, 
like spin symmetries \cite{yl07}.

%In particular, the spectral decomposition of the 
%broken-symmetry GP vortex states in terms of $L$-conserving RVCs
%provides a characteristic fingerprint whose presence or absence in the exact 
%solutions is an indicator of the importance, or not, of vortices in small 
%clouds of rotating bosons.

We derive the RVC wave function by using an adaptation of the two-step 
many-body method of symmetry breaking/symmetry restoration.
We start with the observation that the 
{\it many-body\/} GP vortex solution, $\Psi^{\text{GP}}$, (as well as any 
mean-field solution exhibiting a breaking of the rotational
symmetry) is a wave packet, and thus it can be expanded as a linear 
superposition over eigenstates $\Phi_{N,L}$ of the many-body Hamiltonian 
${\cal H}$ with good total angulal mommentum $L$, i.e.,
\begin{equation}
\Psi^{\text{GP}}_N({\bf r}_1, {\bf r}_2,\ldots,{\bf r}_N) = 
\sum_L C_L \Phi_{N,L} ({\bf r}_1, {\bf r}_2,\ldots,{\bf r}_N).
\label{gpv}
\end{equation}

For the two-dimensional case considered here (which is appropriate for a 
rapidly rotating harmonic trap), the eigenstates $\Phi_{N,L}$'s can be 
approximated by using the projection operator 
\begin{equation}
\hat{\cal P}_L = \frac{1}{2 \pi} 
\int_0^{2 \pi} d\theta e^{i\theta(L-\hat{L})} = \delta(L-\hat{L}),
\label{pro}
\end{equation}
which projects states with good total $L$ out of the GP vortex state.
The RVC wave functions are then given by
\begin{equation}
|\Phi^{\text{RVC}}_{N,L}\rangle=\hat{\cal P}_L |\Psi^{\text{GP}}_N \rangle
= \int_0^{2 \pi} d\theta |\Psi^{\text{GP}}_N(\theta) \rangle e^{i \theta L},
\label{prj}
\end{equation}
where $|\Psi^{\text{GP}}_N(\theta) \rangle$ is the original many-body GP 
vortex solution rotated by an azimuthal angle $\theta$. The projection
techniques use the fact that the broken-rotational-symmetry states form
a manifold of energy degenerate states (i.e., their total energy is 
independent of the azimuthal angle $\theta$); in this respect, the phases
$e^{i \theta L}$ in Eq.\ (\ref{prj}) are the characters of the rotational 
group in two dimensions.

The expansion coefficients $C_L$ in Eq.\ (\ref{gpv}), which specify the 
spectral decomposition of the many-body Gross-Pitaevskii vortex state, can be 
calculated using the projected wave functions (\ref{prj}) for the
$\Phi_{N,L}$'s in the r.h.s of Eq.\ (\ref{gpv}). Taking into consideration that 
$\Psi^{\text{GP}}_N({\bf r}_1, {\bf r}_2,\ldots,{\bf r}_N) = 
\prod_{i=1}^N \phi_0({\bf r}_i)$, one finds
\begin{equation}
C_L = \frac{1}{2 \pi} \int_0^{2 \pi} d\theta n(\theta) e^{i \theta L},
\label{cl}
\end{equation}
where the overlap kernel is given by 
$n(\theta)= \\ \langle \phi_0(\theta=0) | \phi_0(\theta) \rangle^N$,
and the multiply occupied single orbital $\phi_0({\bf r})$ is a solution of 
the familiar Gross-Pitaevskii equation
%\begin{eqnarray}
\begin{equation}
 [ H({\bf r}) + g (N-1) |\phi_0({\bf r})|^2] \phi_0({\bf r})
= \varepsilon_0 \phi_0({\bf r}),
\label{gpe}
\end{equation}
%\end{eqnarray}
with the single-particle Hamiltonian given by 
$H({\bf r}) = {\bf p}^2 /(2m) - \Omega \hat{l} + m \omega_0^2 {\bf r}^2/2$,
where $\Omega$ is the rotational frequency of the trap and
$\omega_0$ characterizes the circular harmonic confinement.

The total energy of the RVC is given by
\begin{equation}
E^{\text{RVC}}_{N,L} = \left. { \int_0^{2\pi} h(\theta) 
e^{i \theta L} d \theta } \right/%
{ \int_0^{2\pi} n(\theta) e^{i \theta L} d \theta},
\label{eproj}
\end{equation}
with the Hamiltonian kernel being 
$h(\theta) = \\ \langle \Psi^{\text{GP}}_N(\theta=0) | {\cal H} | 
\Psi^{\text{GP}}_N(\theta) \rangle$, where the many-body Hamiltonian is 
${\cal H} = \sum_{i=1}^N H({\bf r}_i) + 
g \sum_{i<j}^N \delta({\bf r}_i-{\bf r}_j)$.
The above constitutes an effective continuous configuration-interaction
scheme which lowers the mean-field energy by introducing correlations
\cite{yl07}. The lowering of the ground-state energy brought about by
the angular-momentum projection can be seen (see Ref. \cite{low62}) 
from evaluation of the GP ground-state energy using the spectral
decomposition given in Eq.\ (\ref{gpv}). This yields the
expression $E^{\text{GP}}_N = \sum_L |C_L|^2 E^{\text{RVC}}_{N,L}$    
with $\sum_L |C_L|^2=1$. Since $E^{\text{GP}}_N$ is expressed as
a weighted average of $E^{\text{RVC}}_{N,L}$ with positive
weights, it is obvious that at least one of these energies obeys
$E^{\text{RVC}}_{N,L} \leq E^{\text{GP}}_N$.

%
%****************************** begin figure 1 **************************
\begin{figure}[t]
\centering{\includegraphics[width=6.5cm]{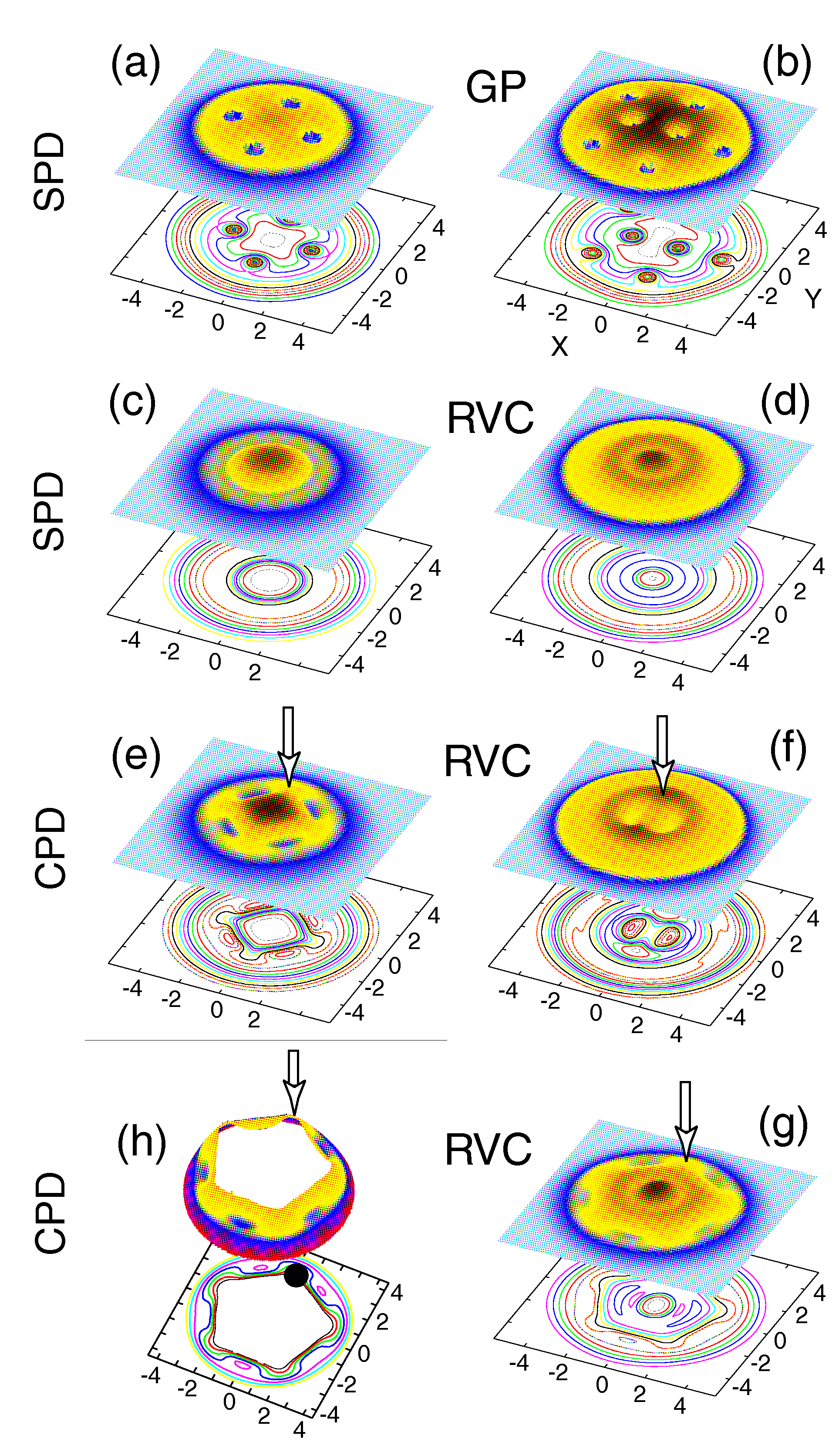}}
\caption{
Rotating vortex cluster and GP vortex solutions for $N=9$ trapped bosons in
a rotating trap with angular frequencies of trap rotation $\Omega/\omega_0=
0.5$ [top three panels in the left column] and $\Omega/\omega_0=
0.565$ [all four panels in the right column plus (h)]. For $\Omega/\omega_0
=0.5$, a $(0,4)$ single polygonal ring of four vortices is involved, while for 
$\Omega/\omega_0=0.565$ a $(2,5)$ 
double polygonal ring of seven vortices develops. (a,b) GP single-particle
densities (SPDs). (c,d) RVC single-particle densities. (e-h) RVC conditional
probability distributions (CPDs), with the fixed point (marked by a thick 
vertical arrow) at ${\bf r}_0=(0,2.07l_0)$ (e), ${\bf r}_0=(0,1.14l_0)$ (f), 
and ${\bf r}_0=(0,3.2l_0)$ (g,h). (h) A horizontal slice of the CPD in (g) which
magnifies the five vortices of the outer ring. The RVC total angular momenta are
$L=28$ [(c) and (e)] and $L=36$ [(d),(f-h)]. The corresponding GP 
total-angular-momentum averages 
($\langle \Psi^{\text{GP}}_N|  \hat{L} | \Psi^{\text{GP}}_N \rangle$) 
are 26.93 and 38.11, respectively.  Note the elliptic shape of the vortex cores
in the CPDs [see (e-h)] reflecting azimuthal fluctuations in the RVC state.
The strength of the interparticle repulsion 
was taken $R_\delta=50$ [$R_\delta \equiv gm/(2\pi \hbar^2)$, see Ref.\ 
\cite{roma04}]. Unit length: $l_0 = \sqrt{ \hbar/(m\omega_0) }$.
}
\label{spdcpds}
\end{figure}
%****************************** end figure 1 **************************

To illustrate the essential qualitative difference between the RVCs and the GP 
vortex states, we contrast in Fig.\ \ref{spdcpds} their single-particle 
densitites (SPDs) for the case of $N=9$ trapped bosons and
when the GP solutions exhibit either a (0,4) single-polygonal-ring of four
[Fig.\ \ref{spdcpds}(a)] or a $(2,5)$ double-polygonal-ring configuration of
seven [Fig.\ \ref{spdcpds}(b)] localized vortices (for the trap and other 
parameters employed, see the caption of Fig.\ \ref{spdcpds}); $(n_1,n_2)$
denotes $n_1$ ($n_2$) vortices on the inner (outer) ring.
In sharp contrast to the GP single-particle densities, 
the RVC SPDs are circularly symmetric: the one [Fig.\ \ref{spdcpds}(c)] 
corresponding (through the aforementioned projection) to the four GP vortices 
exhibits instead a single continuous ring of depleted matter, while the other 
one [Fig.\ \ref{spdcpds}(d)] corresponding to the seven GP vortices exhibits
instead two concentric continuous rings of depleted matter. 

Due to the symmetry restoration, the vortex structures become ``hidden'' in 
the RVC single-particle densities. However, they can be revealed through the 
use of conditional probability distributions (CPDs) defined as
\begin{equation}
P({\bf r},{\bf r}_0) =
\langle \Phi^{\text{RVC}}_{N,L}|
\sum_{i \neq j} \delta({\bf r}_i - {\bf r})\delta({\bf r}_j - {\bf r}_0)
|\Phi^{\text{RVC}}_{N,L} \rangle.
\label{cpd}
\end{equation}
CPDs give the probability of finding a boson at position ${\bf r}$ given
that another boson is located at a fixed point ${\bf r}_0$.

In the CPDs calculated for the RVC states [see Figs.\ \ref{spdcpds}(e,f,g,h)], 
the fixed point is associated with a hump (local maximum) and the vortices are 
given by depressions (local minima) in the matter density. The number of 
vortices of a GP state and of an RVC projected out from it is the same for all
$\Omega$'s; e.g., in the CPD in Fig.\ \ref{spdcpds}(e) four vortices are seen 
corresponding to the four GP vortices in Fig.\ \ref{spdcpds}(a). This reflects 
the fact that the projection maintains the same (intrinsic or hidden) point-group
symmetry.

For the $(2,5)$ double-ring RVC, the fixed point can be placed on the inner or
the outer ring. In the first case, the calculated CPD [Fig.\ \ref{spdcpds}(f)] 
shows two vortices on the inner ring and remains
uniform along the outer ring, while in the second case
the calculated CPD [Fig.\ \ref{spdcpds}(g,h)] shows five vortices on 
the outer ring and remains uniform along the inner ring. This suggests that
the rings rotate independently of each other in analogy with the case of 
rotating boson molecules \cite{roma06} and rotating electron molecules in 
quantum dots \cite{yl07}. 
%
%****************************** begin figure 2 **************************
\begin{figure}[t]
\centering{\includegraphics[width=8.0cm]{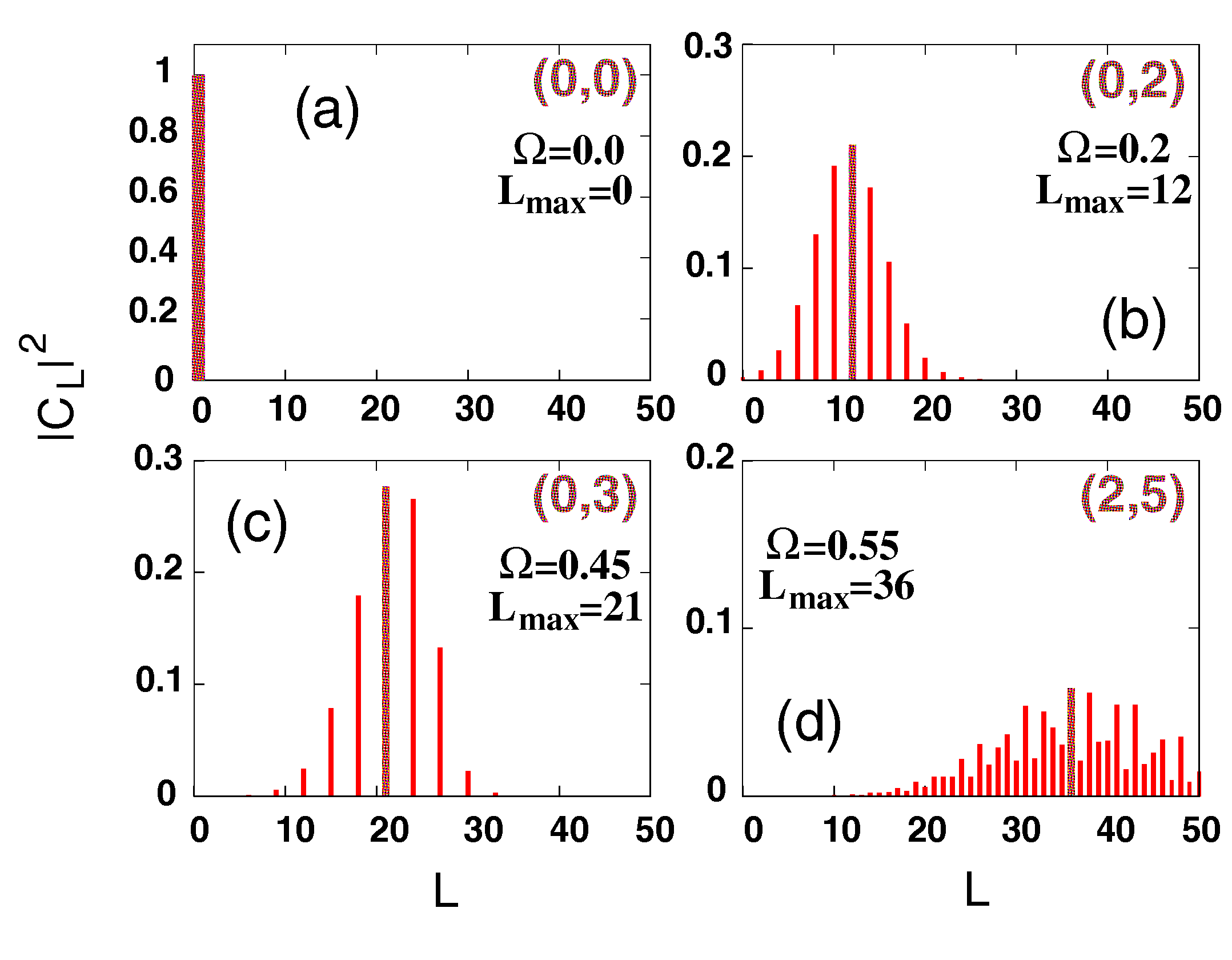}}
\caption{
Spectral decomposition [$C_L$ coeffiecients modulus square, Eq.\ 
(\ref{cl})] for GP vortex ground states in characteristic cases.
Angular frequency of the trap: (a) $\Omega/\omega_0=0$. 
(b) $\Omega/\omega_0=0.2$. (c) $\Omega/\omega_0=0.45$.
(d) $\Omega/\omega_0=0.55$. The total angular momenta associated with 
the largest coefficients are $L_{max}=0$, 12, 21, and 36, respectively. The
polygonal ring configurations of the GP vortices are also marked as $(n_1,n_2)$.
The ratio of the interparticle repulsion and the kinetic energy was taken 
$R_\delta=50$ (see caption of Fig. \ref{spdcpds}). In the figure, $\Omega$ is 
given in units of $\omega_0$. The GP total-angular-momentum expectation values 
are (a) 0.0, (b) 11.01, (c) 21.42, and (d) 36.53.
}
\label{gpspdec}
\end{figure}
%****************************** end figure 2 **************************

To further investigate the wave-packet properties of the GP vortices, we have
numerically calculated their spectral decomposition in terms of RVC states [see 
Eq.\ (\ref{gpv})]. In Fig.\ \ref{gpspdec}, we plot the expansion coefficients 
$C_L$ [calculated numerically according to Eq.\ (\ref{cl})] for $N=9$ bosons
and for several characteristic angular frequencies. The $\Omega=0$ case is a
rather trivial one where the GP solution preserves the circular symmetry, 
exhibits no vortices, and coincides with the corresponding RVC for $L=0$ (in 
this limiting case, one has $C_0=1$ and $C_L=0$ for any $L \ne 0$). 

For the chosen value $R_\delta=50$ (see the caption of Fig.\ \ref{spdcpds}), 
non-trivial cases arise for $\Omega/\omega_0> 0.175$. 
For $\Omega/\omega_0=0.2$
[Fig.\ \ref{gpspdec}(b)] and $\Omega/\omega_0=0.45$ [Fig.\ \ref{gpspdec}(c)]
the GP vortex solutions exhibit a (0,2) and (0,3) single polygonal-ring
configurations, respectively. The expansion
coefficients $C_L$ clearly demonstrate that the GP vortex states given as
examples in these figures can be 
reconstructed from linear superpositions of RVC states [observe the many 
non-vanishing values of $C_L$ in Fig.\ \ref{gpspdec}(b) and 
Fig.\ \ref{gpspdec}(c)]. Of central importance is the selection rules that
the RVC angular momenta must obey in order to be compatible with the $(n_1,n_2)$ 
intrinsic RVC point-group symmetry (which coincides with the explicit
point-group symmetry of the associated GP vortex states). Indeed, for the (0,2) 
vortex ring the RVC angular momenta obey the relation $L=2k$, while for the 
(0,3) case, one has $L=3k$, with $k=0,1,2,...$. The RVC angular momenta 
vary in a stepwise manner, and the value of the step coincides with
the number of GP vortices on the single ring.
For $\Omega/\omega_0=0.55$, the GP vortex state exhibits a double
(2,5) polygonal-ring structure, which imposes upon the RVC decomposition a more 
complex, but approximate, selection rule $L=2k_1+5k_2$, with both $k_1$ and
$k_2$ being positive integer numbers. While the larger $C_L$'s 
[Fig.\ \ref{gpspdec}(d)] conform to this rule, there are several
smaller $C_L$'s whose angular momenta fall outside this rule. The reason is
that in the GP solution [see Fig.\ \ref{spdcpds}(b)] the arrangement of 
vortices on the outer ring deviates slightly from being a perfect regular 
pentagon \cite{note2}.

From the many RVC states that contribute at a given $\Omega$ to the spectral 
decomposition of the GP vortex states (including in the latter both ground and
low-lying excited configurations), there is one with lowest energy 
$E^{\text{RVC}}_{\text{GS}}$, 
which is the ground state within the RVC approximation at this specific
rotational frequency. For the $\Omega$'s and the parameters considered in Fig.\
\ref{gpspdec}, we found that the RVC ground states have angular momenta 
$L_{\text{GS}}$ associated with the largest coefficients $|C_L|$ in the
GP decompositions; these $L_{\text{GS}}$'s always obey the polygonal-ring 
selection rules discussed above. 

\begin{figure}[t]
\centering{\includegraphics[width=7.50cm]{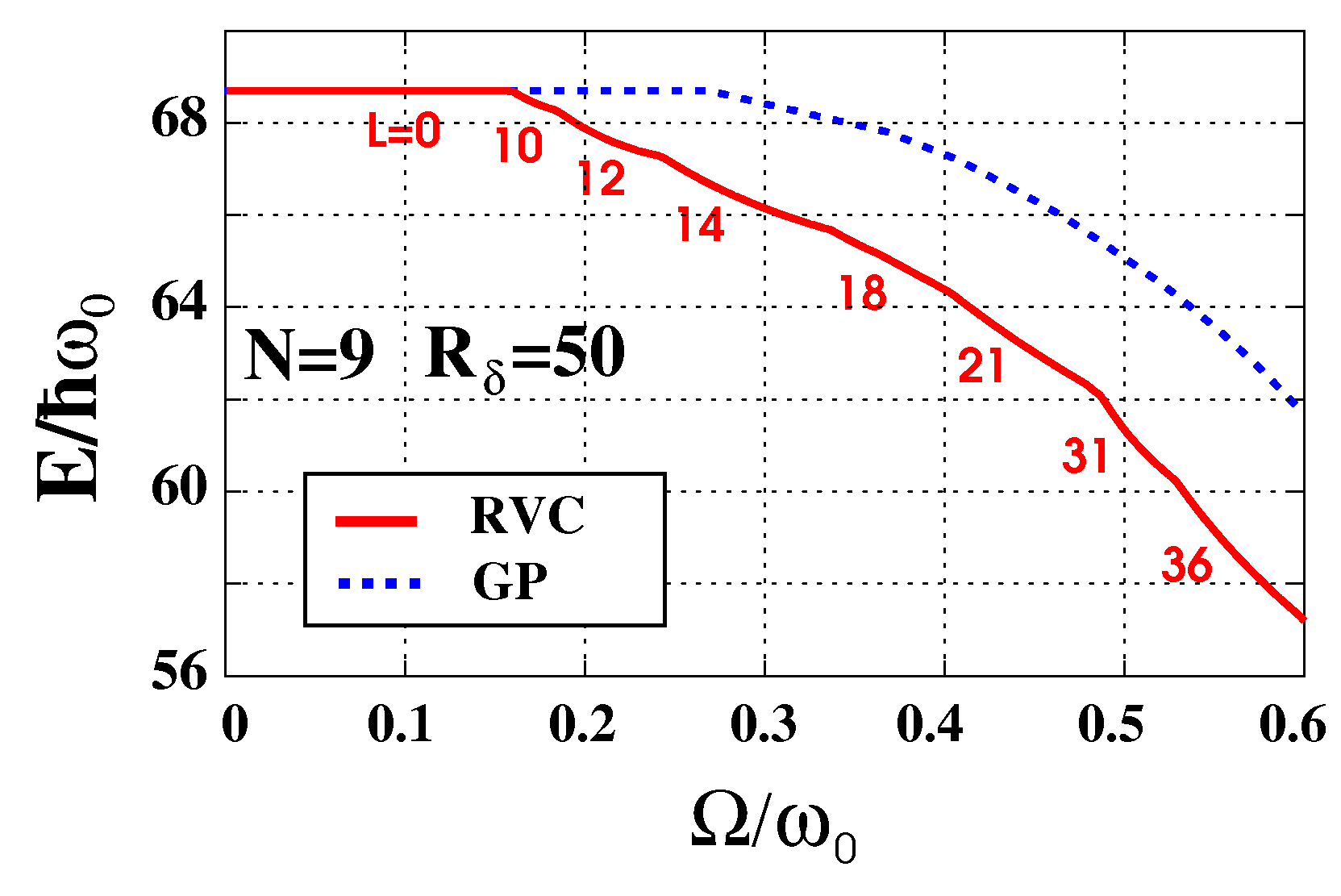}}
\caption{
RVC (solid) and GP (dotted) ground-state energies for $N=9$ bosons and 
$R_\delta=50$, plotted versus the rotational frequency $\Omega/\omega_0$.
The ground-state angular momenta are denoted under the RVC curve.
}
\label{rvcremen}
\end{figure}

\begin{figure}[t]
\centering{\includegraphics[width=7.50cm]{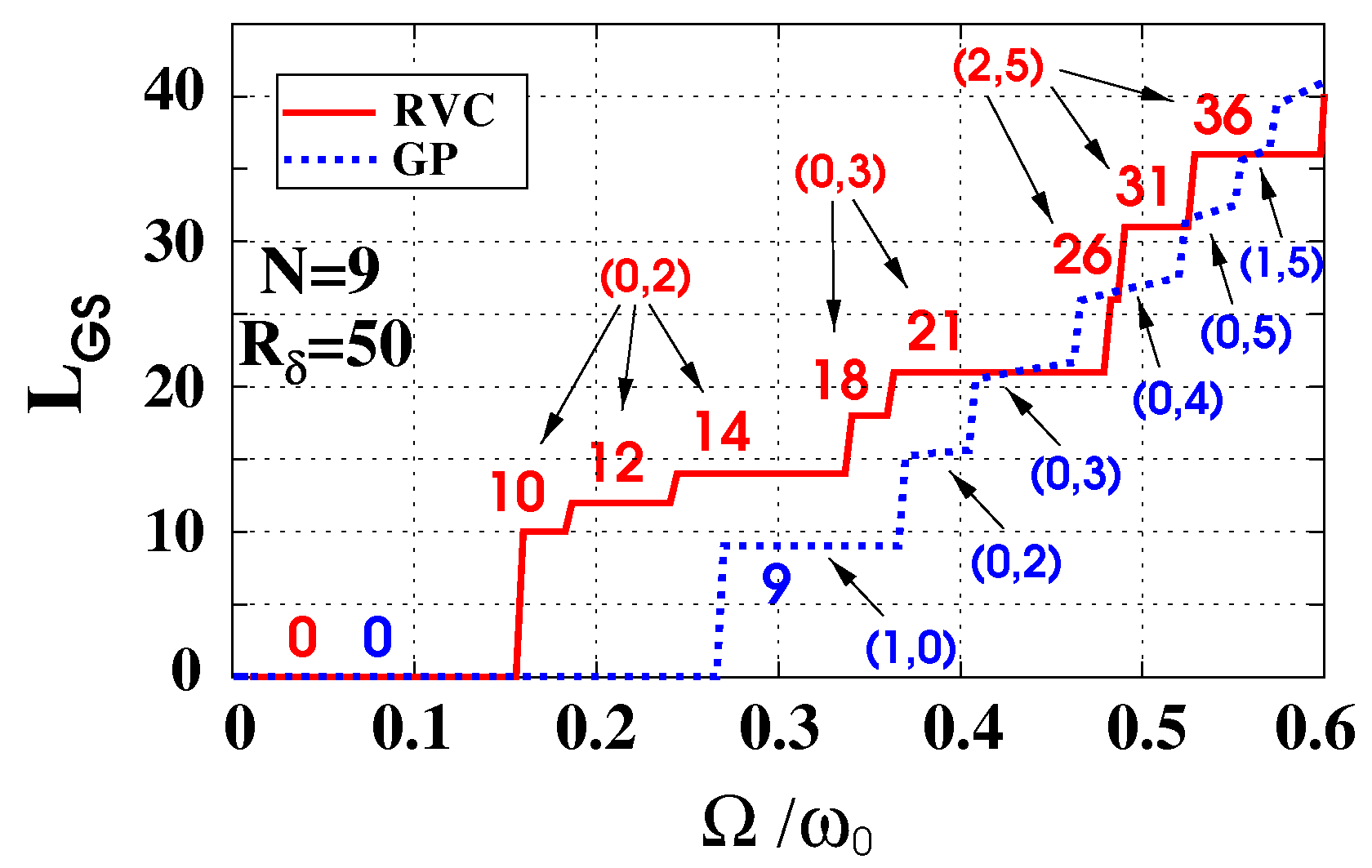}}
\caption{
RVC (solid) and GP (dotted) ground-state angular momenta for $N=9$ bosons and
$R_\delta=50$, plotted versus the rotational frequency $\Omega/\omega_0$.
The quantized RVC angular momenta are explicitly denoted in the figure, along
with the corresponding polygonal-ring configurations $(n_1,n_2)$. The 
Gross-Pitaevskii $(n_1,n_2)$ configurations are also denoted. The GP
angular momenta are not quantized; they are given by the expectation 
values $\langle \Psi^{\text{GP}}_N|  \hat{L} | \Psi^{\text{GP}}_N \rangle$.
}
\label{gsangmom}
\end{figure}

As aforementioned from the general theory of projection techniques \cite{low62}, the 
RVC ground-state energies are lower than (or equal at most to) the corresponding
GP ones for all values of the rotational frequency (Fig.\ \ref{rvcremen}). 
Furthermore, the angular momenta associated with the RVC ground states are quantized
and thus exhibit a step-wise increase as a function of the rotational frequency (see
Fig.\ \ref{gsangmom}). The magnitude of the steps in 
the ground-state RVC angular momenta changes 
with $\Omega$, since RVCs with different $(n_1,n_2)$ intrinsic vortex configurations
become the ground-state as $\Omega$ increases. This behavior contrasts with that of 
the ground-state GP angular momenta that vary continuously as a function of $\Omega$
without any direct association to the $(n_1,n_2)$ vortex configuration
(Fig.\ \ref{gsangmom}).
In several instances, the RVC ground state has an intrinsic $(n_1,n_2)$
point-group symmetry that is different from that of the GP vortex ground state
at the same $\Omega$. This happens when the projection of an excited GP state
results in a larger energy gain compared with that of the GP ground state.

%
%****************************** begin figure 3 **************************
%****************************** end figure 3 **************************

While we focus here on obtaining a proper symmetry-conserving
vortex theory for a finite (small) number of trapped bosons, we comment on
the RVC behavior compared with that obtained through exact diagonalization
(EXD) calculations, which have become in recent years computationally feasible 
for smaller $N$. In particular, unlike the RVC case studied here (see Fig.\
\ref{gsangmom}), EXD calculations in the lowest Landau 
level for $N=9$ bosons exhibit quantized ground-state ``magic'' angular momenta
that follow a $L_m=n_1k_1+n_2k_2$ selection rule, but with the additional
condition $n_1+n_2=N$ (unlike the RVC behavior where $n_1+n_2=q \neq N$); 
see, e.g., Fig.\ 10 in Ref.\ \cite{baks07} and Fig.\ 2 in Ref.\ \cite{barb06}. 
Moreover, it was shown in Ref.\ \cite{baks07} that  
EXD ground-state wave functions describe formation of a rotating boson molecule 
\cite{roma04,roma06} exhibiting two distinct (1,8) and (2,7) isomers of
localized bosons. 

The above characteristic difference between the RVC and RBM 
\cite{roma06,baks07} solutions maintains for other values of $N$ and $R_\delta$,
reflecting the intrinsic point group symmetry of 
the angular momentum conserving theory 
(RVC, RBM via projection methods \cite{roma06} or EXD \cite{baks07}).  This is 
particularly the case for low $N$ and high $R_\delta$, where the RBMs are 
energetically favored and the RVCs may be regarded as higher lying excited 
states. However, for higher $N$ (occurring earlier for low $R_\delta$) the RVC 
may compete effectively with the localized RBM states, and eventually become the 
ground state.  

%
%****************************** begin figure 4 **************************
%****************************** end figure 4 **************************

In conclusion, we have introduced a correlated many-body wave function,
referred to as a rotating vortex cluster, which conserves the total
angular momentum and has lower energy compared to the Gross-Pitaevskii solution.
The RVC is better suited to describe formation of vortices in 
small rotating clouds of trapped bosons compared to the mean-field 
GP vortex states that break the rotational symmetry. The 
GP vortex states were shown to be wave packets composed of such RVC states.
The calculation of the properties of rotating-vortex-cluster states allowed
for comparisons of qualitative signatures (e.g., ground-state angular momenta 
sequences) between the RVC and exact-diagonalization results. We conclude that
the physics of small rotating bosonic clouds is markedly different from
that of larger assemblies known to behave as ideal Bose-Einstein condensates
(properly described by the broken symmetry GP vortex solutions).
We hope that these results will motivate further experimental research 
in the area of correlated states in small bosonic systems (see, e.g., Ref.\ 
\cite{geme}).

This work was supported by the US D.O.E. (Grant No. FG05-86ER45234).


\begin{thebibliography}{99}
\bibitem{note2}
The five GP vortices [Fig.\ \ref{spdcpds}(b)] 
are located at different distances from the
center at $3.29 l_0$, $2.78 l_0$, $2.96 l_0$, $2.96 l_0$, and $2.78 l_0$
(the azimuthal angles between neighboring vortices also deviate slightly from
the ideal value of $2\pi/5$). This originates from intervortex repulsion.
\end{thebibliography}
\end{document}